\documentclass[10pt,journal,compsoc]{IEEEtran}
\usepackage{amsmath,amssymb,amsfonts}
\usepackage{algorithmic}
\usepackage{graphicx}
\usepackage{textcomp}
\usepackage{xcolor}
\usepackage{bbding}
\usepackage{subfigure}
\usepackage{enumitem}
\usepackage{multirow}
\usepackage{booktabs}
\usepackage{color}
\usepackage{amsthm}
\usepackage{pgfplots}
\usepackage{ulem}
\usepackage{flushend}

\usepackage{bm}
\usepackage{url}
\usepackage{makecell}

\newcommand{\Mat}[1]{\bm{#1}}
\newcommand{\Vector}[1]{\bm{#1}}
\newcommand{\Set}[1]{\mathcal{#1}}

\newcommand{\Loss}{\mathcal{L}}

\ifCLASSOPTIONcompsoc
  \usepackage[nocompress]{cite}
\else
  \usepackage{cite}
\fi
\ifCLASSINFOpdf
\else
\fi
\begin{document}

%
\title{Knowledge-aware Collaborative Filtering with Pre-trained Language Model for Personalized Review-based Rating Prediction}
%
%
%
%

\author{Quanxiu Wang, Xinlei Cao, Jianyong~Wang,~\IEEEmembership{Fellow,~IEEE,} Wei~Zhang,~\IEEEmembership{Member,~IEEE}
\IEEEcompsocitemizethanks{\IEEEcompsocthanksitem Q. Wang, X. Cao, and W. Zhang (corresponding author) are with the School of Computer Science and Technology, Shanghai Institute for AI Education, East China Normal University, Shanghai 200062, China (e-mail:51205901126@stu.ecnu.edu.cn, cyril.xlcao@gmail.com, zhangwei.thu2011@gmail.com). W. Zhang is also with KLATASDS-MOE.

\IEEEcompsocthanksitem J. Wang is with the Department of Computer Science and Technology, Tsinghua University, Beijing 100086, China, and also with the Jiangsu Collaborative Innovation Center for Language Ability, Jiangsu Normal University, Xuzhou 221009, China (e-mail: jianyong@tsinghua.edu.cn).

\IEEEcompsocthanksitem This paper was partially supported by the National Natural Science Foundation of China (No. 62072182, No. 92270119), and the Fundamental Research Funds for the Central Universities.
}

\thanks{Manuscript received xxxx, xxxx; revised xxxx, xxxx.}}

%
%

\markboth{Journal of \LaTeX\ Class Files,~Vol.~xx, No.~x, August~xxxx}%
{Shell \MakeLowercase{\textit{et al.}}: Bare Demo of IEEEtran.cls for Computer Society Journals}
%



\IEEEtitleabstractindextext{%
\begin{abstract}
Personalized review-based rating prediction aims at leveraging existing reviews to model user interests and item characteristics for rating prediction.
Most of the existing studies mainly encounter two issues.
First, the rich knowledge contained in the fine-grained aspects of each review and the knowledge graph is rarely considered to complement the pure text for better modeling user-item interactions.
Second, the power of pre-trained language models is not carefully studied for personalized review-based rating prediction.
To address these issues, we propose an approach named Knowledge-aware Collaborative Filtering with Pre-trained Language Model (KCF-PLM).
For the first issue, to utilize rich knowledge, KCF-PLM develops a transformer network to model the interactions of the extracted aspects w.r.t. a user-item pair.
For the second issue, to better represent users and items, KCF-PLM takes all the historical reviews of a user or an item as input to pre-trained language models.
Moreover, KCF-PLM integrates the transformer network and the pre-trained language models through representation propagation on the knowledge graph and user-item guided attention of the aspect representations.
Thus KCF-PLM combines review text, aspect, knowledge graph, and pre-trained language models together for review-based rating prediction.
We conduct comprehensive experiments on several public datasets, demonstrating the effectiveness of KCF-PLM.
\end{abstract}

}

\maketitle

\IEEEdisplaynontitleabstractindextext

%
\IEEEpeerreviewmaketitle


%
%
%
%

\section{Introduction}\label{sec:intro}

\IEEEPARstart{I}n the information explosion era, ordinary users not only have the identity as consumers, but also play a role as content producers.
The effect has proved to be a double-edged sword.
On the one hand, it becomes more difficult for users to find what they are interested in due to the sheer volume of big data.
On the other hand, characterizing user interests with their own generated data is more promising.
User reviews are one kind of representative user-generated data, which are often seen in e-commerce platforms.
Typically, each review involves a user who generates it, an item being rated, a rating score, and review text.
Many research efforts have been devoted to utilizing reviews for personalized rating prediction~\cite{MF} that estimates missing rating scores for different user-item pairs.
Compared with conventional rating prediction that mainly relies on users, items, and rating scores, personalized review-based rating prediction~\cite{McAuley-RecSys13} additionally leverages review text for achieving better prediction performance.

\begin{figure*}[!t]
\centering
\includegraphics[width=.95\linewidth]{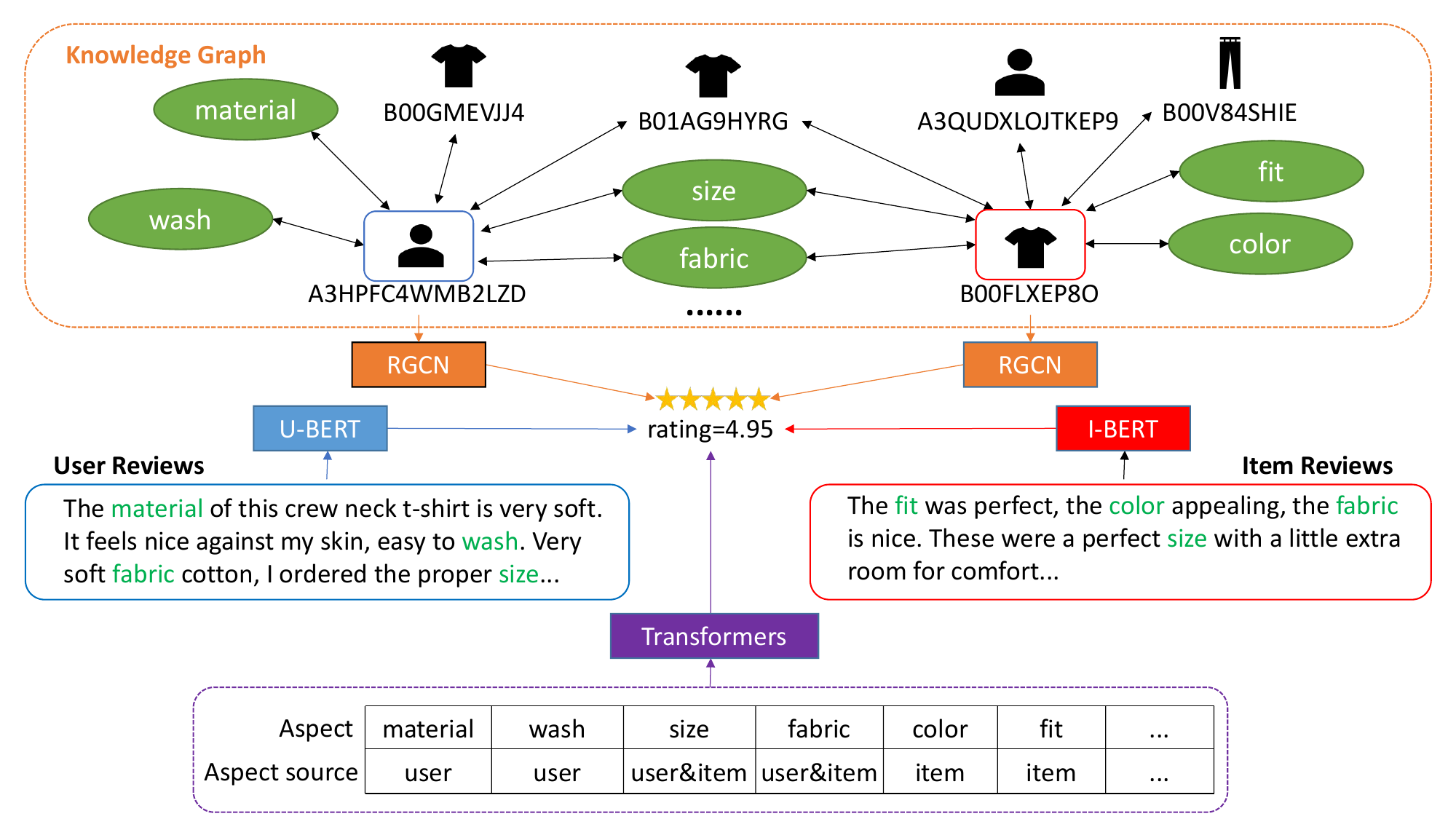}
\caption{Conceptual diagram of how KCF-PLM works for a running case corresponding to a sampled user-item pair in E-commerce.} \label{fig:running-case}
\end{figure*}

Early personalized review-based rating prediction methods are keen on topic models to learn user or item representations ~\cite{McAuley-RecSys13,BaoFZ14,itlfm}.
These representations enhance the expressive ability of latent factor models such as matrix factorization~\cite{MF} towards calculating ratings.
Later on, deep learning-based models~\cite{deepconn,narre,daml,SSG} become the mainstream approaches for this task.
Due to the strength of performing deep representation learning directly on the unstructured raw text of reviews, they achieve improved performance over the early methods for rating prediction.

Although much progress has been made to push forward research development and industrial application, most of the existing methods suffer from two issues.
First, there lacks an effective manner of integrating the modeling of unstructured text and fine-grained aspects.
In reality, both the pure unstructured review text and its involved fine-grained aspects contain a wealth of knowledge.
For example, as a real case shown in Figure~\ref{fig:running-case}, the aspects, marked by green in the review text, address the key attributes of items and could reflect the attention of users.
Besides these aspects, other words such as `soft' and `nice' depict the user's attitude towards these aspects.
Moreover, the introduction `with a little extra room for comfort' also indicates the sentiment polarity to a certain extent.
Without modeling the pure unstructured text, the useful knowledge besides aspects might not be captured by rating prediction methods.
Existing studies either focus on modeling pure unstructured text~\cite{deepconn,mpcn,daml} or considering mining aspects from review text and learning aspect representations for rating prediction~\cite{ZhangL0ZLM14,AARM,LiQPQDW19,ZhaoLXDS20}.
However, few studies consider simultaneously utilizing them in a unified model for personalized review-based rating prediction.
Motivated by this, a natural research question arises: how to effectively model the fine-grained aspects and unstructured review text to make them complement each other for the studied task.
Second, pre-trained language models (PLMs)~\cite{bert} are rarely considered for the personalized review-based rating task, although they exhibit superior performance in various downstream tasks.
To our knowledge, there is one pre-trained language model study called NCEM~\cite{NCEM}.
But it only utilizes the pre-trained language model BERT~\cite{bert} separately from the rating prediction model, without fine-tuning it.
Therefore, another research question is how to effectively incorporate pre-trained language models into the rating prediction approach.

To answer the above two research questions, in this paper, we propose Knowledge-aware Collaborative Filtering with Pre-trained Language Model (KCF-PLM), a novel model that combines review text, aspect, knowledge graph (KG), and pre-trained language model for personalized review-based rating prediction.
For the first question, we extract all the aspects from raw reviews and construct a KG wherein the nodes are composed of users, items, and aspects.
Then for a given user-item pair to be predicted, we collect the aspects extracted from the reviews of the user and the item.
We define a field named aspect source to distinguish whether an aspect comes from the user, the item, or even both sides.
Figure~\ref{fig:running-case} shows an example, where `size' occurs in both the user reviews and the item reviews, while `material' only occurs in the user reviews.
KCF-PLM leverages a transformer network to model the interactions of the aspects w.r.t. the user-item pair.
As such, the aspect representations could be contextualized, and the matching of user attention and item attributes could be captured.

For the second question, KCF-PLM utilizes two pre-trained language models (e.g., BERT~\cite{bert}) to learn user and item representations based on their review text, respectively.
The pre-trained language models are incorporated into the whole rating prediction model by the following two manners. (1) Representation propagation on the KG. 
KCF-PLM uses graph neural networks (e.g., RGCN~\cite{rgcn}) to propagate representations so that different types of node representations are mutually affected.
For example, the representation of the user `A3HFC4WMB2LZD' is affected by the aspect `material' and the item `B00GMEVJJ4' in the figure. 
The representations of user and item nodes are then combined with the user and item representations from the pre-trained language models.
And the representations of aspect nodes are also used to enhance the transformer network.
(2) Personalized attention-based combination of the aspects.
KCF-PLM further utilizes the combined user and item representations to guide the attention-based combination of the aspect representations from the transformer network.
This is motivated by the fact that the importance of an aspect varies from one user-item aspect to another user-item aspect.

In summary, the main contributions of this paper are as follows:
\begin{itemize}
\item We propose KCF-PLM to model review text, aspect, and KG in a unified architecture, and thus rich knowledge could be captured. To the best of our knowledge, this is the first model that simultaneously considers these factors for the task of personalized review-based rating prediction.

\item We develop a transformer network to learn the interactions of structured review aspects and leverage pre-trained language models to learn the representations of unstructured review text in an end-to-end fashion. 

\item We conduct experiments on several real public datasets, demonstrating the proposed KCF-PLM significantly outperforms strong competitors and validating the effectiveness of the main considerations in KCF-PLM. Our source code is available at \url{https://github.com/Wqxleo/KCF-PLM}.
\end{itemize}

\section{Related Work}
In this section, we review the related studies from the following three aspects: personalized review-based rating prediction, knowledge graph-based recommendation, and pre-trained language models for recommendation.

\subsection{Personalized Review-based Rating Prediction}\label{subsec:prrr}
Personalized rating prediction and top-N recommendation are the two main task types in recommender systems~\cite{ShiLH14}.
The contrast is that the former one is a regression task~\cite{PoudelB22} while the latter one is a ranking task.
This paper concentrates on personalized review-based rating prediction.
Compared with conventional personalized rating prediction, personalized review-based rating prediction~\cite{McAuleyL13} considers not only user, item, and rating, but also review text information.
In the last decade, many efforts have been devoted to this task.
Basically, the rating prediction models could be summarized into two categories: shallow rating prediction~\cite{McAuleyL13,BaoFZ14,itlfm,ZhangYHW16} and deep rating prediction~\cite{deepconn,narre,daml,SSG}.

The pioneering study HFT~\cite{McAuleyL13} in shallow rating prediction utilizes probabilistic topic models~\cite{Blei03-JMLR} to learn a topic distribution for each review.
The representation of either the user or the item associated with the review is directly transformed by the topic distribution.
TopicMF~\cite{BaoFZ14} makes a step further by simultaneously associating both user and item representations with review topic distributions.
To utilize both advantages of latent factors learned by matrix factorization and topic factors learned by topic models, ITLFM~\cite{itlfm} combines them to form unified representations and exhibits better performance than HFT and TopicMF.
However, these shallow rating prediction models have a limited model capacity and thus it is hard for them to learn more expressive representations.

Thanks to the wave of development in deep learning~\cite{lecun2015deep}, deep learning-based rating prediction models have emerged and dominated this field.  
The major reason is attributed to the fact that deep learning does well in representation learning, which is crucial for review text modeling.
DeepCoNN~\cite{deepconn} is a representative study that uses convolutional neural networks for the task.
For a given user-item pair, DeepCoNN fuses all the reviews of the user into a long review document and this operation is repeated for the item.
Then the convolutional neural network builds the user representation and the item representations based on the corresponding long review document, respectively.
On top of the convolutional neural networks, the user and item representation are concatenated and fed into Factorization Machine (FM)~\cite{Rendle12} to estimate the rating.
Later on, NARRE~\cite{narre} argues that different reviews of a user or an item do not have exactly the same importance.
Attention mechanism~\cite{BahdanauCB14} is introduced to compute the importance weights.
They are used to combine review representations for representing the user or the item. 
Similarly, CARL~\cite{WuQLWZL19} exploits the attention mechanism to calculate the importance of each word to represent users and items.
In summary, all the above models learn user review text and item review text in a separate manner, without capturing their mutual interactions.

To alleviate this issue, the models such as MPCN~\cite{mpcn}, DAML~\cite{daml}, and AHN~\cite{ahn} are proposed.
For MPCN, its most intriguing part is the review-level co-attention which calculates the affinity between anyone review from a user and anyone review from an item.
The affinity could be further used to select the most important review for composing user and item representations.
Similarly, DAML performs mutual attention between the local contextual features of a user and an item to capture the correlation of their reviews.
The local contextual features are obtained by convolution operation over the word vector matrices of reviews.
As such, both MPCN and DAML realize mutual attention between users and items from a symmetrical perspective.
The main difference between MPCN and DAML is the semantic level considered for attention computation.
On the contrary, AHN implements co-attention with an asymmetrical perspective.
It means that the attention-based review aggregation on the user side is guided by the item side, while on the item side, the review aggregation is not affected by the user side.

To sum up, all the above models concentrate on directly modeling unstructured review text.
Besides these models, there are some other studies investigating personalized review-based rating prediction from new perspectives.
For example, the study~\cite{LyuY0LLD21} considers the reliability of reviews by arguing that fake reviews would hurt the modeling performance.
It formulates a multi-task learning setting to simultaneously predict the rating score and the reliability score.
In contrast, this paper focuses on not only modeling the unstructured review text but also leveraging the fine-grained aspects extracted from reviews to achieve personalized review-based rating prediction.

To our knowledge, there are a few studies that consider modeling the aspects of the studied task.
Specifically, EFM~\cite{ZhangL0ZLM14} incorporates explicit factors related to aspects into matrix factorization.
Since it mainly focuses on explainability, there is much room for improving its rating prediction performance.
ANR~\cite{ChinZJC18} builds aspect-level user and item representations, and the importance of each aspect is estimated to generate overall ratings.
AARM~\cite{AARM} performs interaction modeling on the combinatorial pairs of the aspects, wherein each pair is composed of one aspect from the user side and one aspect from the item side.
CARP~\cite{LiQPQDW19} associates each aspect representation with a sentiment capsule for rating prediction, which is beneficial for understanding the attitude of a user towards an aspect. 
Besides, aspect matching between different domains is also utilized for cross-domain rating prediction~\cite{ZhaoLXDS20}.
However, these studies are only based on aspect modeling, overlooking simultaneously learning from both unstructured text and structured aspects.

\subsection{Knowledge Graph-based Recommendation}
Knowledge graph usually contains abundant relations between various entities and sees many applications in recommender systems~\cite{Guo2020-survey,ShiHSWWDY21}.
Basically speaking, the users and items occurring in recommender systems can be linked to the nodes in a knowledge graph.
Then the user-item interaction modeling could be conducted accompanied by knowledge graph representation learning.
As such, the rich entity and relation information contained in a knowledge graph helps to improve the precision of recommendation. 

Specifically, DKN~\cite{WangZXG18}, proposed for news recommendation, adopts knowledge graph embeddings methods (e.g., TransE~\cite{BordesUGWY13}) to learn entity representations.
Afterward, knowledge-aware CNN is devised to perform convolution operations on the multi-channel input composed of word embeddings, transformed entity embeddings, and transformed context embeddings.
To better unify the knowledge graph learning and recommendation, KTUP~\cite{Cao-WWW2019} uses TransH~\cite{WangZFC14} and hyperplane-based translation for generating recommendations.
Compared to the graph embedding methods used in the above studies, graph neural networks~\cite{gcn,HamiltonYL17} model the high-order propagation among nodes and have also been found in knowledge graph-based recommendation~\cite{WangHWYL0C21}.
Some recent studies~\cite{WangWX00C19,XianFMMZ19,GengFTGMZ22} try to utilize attention mechanisms or reinforcement learning for reasoning in knowledge graphs, which could provide explanations towards recommendations.

However, for the personalized review-based rating prediction task, there are only a few studies~\cite{WuWQGHX19,SSG} that exploit knowledge graphs. 
But only considering user and item nodes is relatively limited.
Sometimes these graphs are regarded as bipartite graphs.
Contrary to these studies, the knowledge graph constructed in this paper contains fine-grained aspects.
As such, richer knowledge is promising to be distilled.

\subsection{Pre-trained Language Models for Recommendation}
In recent years, pre-trained language models (e.g., BERT~\cite{bert} and Transformer-XL~\cite{DaiYYCLS19}) have achieved great success in natural language processing and have been applied to other fields.
The intuition behind the success is that pre-training is conducted on large-scale data in an unsupervised manner and fine-tuning on downstream tasks enables inheriting rich knowledge learned in the pre-training stage.
In the domain of recommender systems, PLMs are usually applied to tasks that require sequential modeling.
For example, in news recommendation, NewsBERT~\cite{WuWYQHL21} leverages BERT to encode the representation of each news.
These news representations then form the user embedding and candidate news embedding.
CTLE~\cite{LinW0L21} considers the characteristics of the user's next location recommendation by introducing another pre-training loss beyond the commonly-used masked language model objective.
This loss regards temporal information as another prediction target.
PeterRec~\cite{Yuan0KZ20} utilizes a convolutional network instead of transformer networks as backbones to model user sequential behaviors.

In this paper, we aim at exploiting PLMs to encode review text representations for facilitating rating prediction.
There are only a few existing studies that consider using PLMs for personalized review-based rating prediction.
Although NCEM~\cite{NCEM} uses BERT, it is used separately from the rating prediction model and fine-tuning is not realized.
In contrast, KCF-PLM incorporates PLMs into the rating prediction model and jointly trains them to predict ratings.

\section{Problem Formulation}\label{sec:problem}
\begin{figure*}[!t] 
\centering
\includegraphics[width=0.85\linewidth]{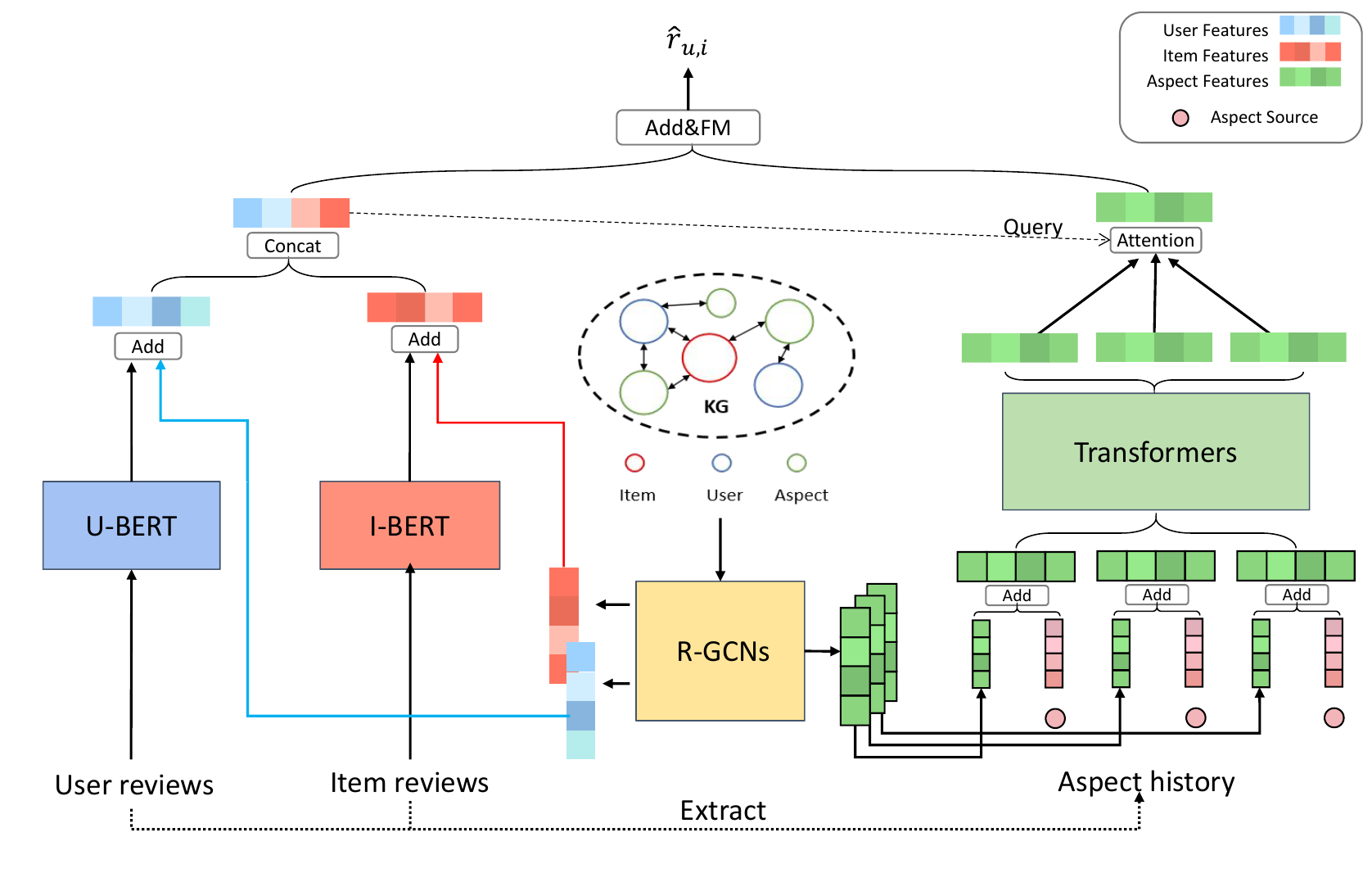}
\caption{Overview of the proposed KCF-PLM. In the figure, the arrow indicates the data flow from input to output, `Add' means the sum of two or more vectors, `Concat' represents the concatenation of vectors, `Attention' means the user-item guided attention in Section~\ref{sec:aspect}, and FM is short for factorization machine.}
\label{fig:model}
\end{figure*}
We begin this section with some basic notations.
Suppose that we are given a user set $\Set{U}=\{u_1,u_2,\cdots,u_{|\Set{U}|}\}$ and an item set $\Set{I}=\{i_i,i_2,\cdots,i_{|\Set{I}|}\}$, where $|\Set{U}|$ is the size of the user set and $|\Set{I}|$ is the size of the item set.
The users and items form a user-item rating matrix $\Mat{Y} = \{y_{ui}\}_{|\Set{U}|\times |\Set{I}|}$, where $y_{ui}$ is the rating score user $u$ gives to item $i$.
In reality, only a part of the entries in the matrix exist, and many other entries are unknown.
In personalized rating prediction, the scores of these unknown entries require to be estimated.
For the available entries, there is review text information besides rating scores.
For example, there is $\Set{D}_{ui} = \{w_1, w_2, \dots, w_{\ell_{ui}}\}$ for rating $y_{ui}$, where $\ell_{ui}$ is the length of the review text.
Based on this, we can combine all the available review text of user $u$ to form a document denoted as $\Set{D}_{u} = \{w_1, w_2, \dots, w_{\ell_{u}}\}$.
And it is similar for item $i$ to have $\Set{D}_{i}$.

Since the aspects of reviews are considered for rating prediction, we suppose the extracted aspects of $\Set{D}_{ui}$ are denoted as $\Set{A}_{ui}=\{a_1,a_2,\cdots, a_{s_{ui}}\}$, where $s_{ui}$ is the size of the aspect set.
Analogously, we have $\Set{A}_{u}$ and $\Set{A}_{i}$ for user $u$ and item $i$, respectively.
Based on the aspects, we can build a knowledge graph $\boldsymbol{G}$ which mainly contains user, item, and aspect as nodes.
The details of the knowledge graph will be introduced later.
Then we formulate the studied problem of this paper as follows.

\newtheorem*{problem}{Problem}
\begin{problem}[Personalized Review-based Rating Prediction] 
Given a target user-item pair with user $u$ and item $i$, the goal of this problem is to learn a model that can: (1) learn from the user reviews $\Set{D}_{u}$, the item reviews $\Set{D}_{i}$, the user aspects $\Set{A}_{u}$, the item aspects $\Set{A}_{i}$, and the knowledge graph $\boldsymbol{G}$, (2) output the estimation of the unknown rating score $\hat{y}_{ui}$, defined as
\begin{equation*}
    \hat{y}_{ui}=f(u, i, \Set{D}_{u}, \Set{D}_{i}, \Set{A}_{u}, \Set{A}_{i}, \boldsymbol{G})\,.
\end{equation*}
\end{problem}

In the following of this paper, matrices, vectors, and scalars are denoted by bold upper case letters, bold lower case letters, and normal lower case letters.

\section{The Proposed Model}\label{sec:model}
In this section, we introduce the proposed KCF-PLM in detail.
First, an overview of the model is illustrated.
Subsequently, we clarify the procedures of aspect extraction and KG construction.
Then the two key modules of PLMs for representation learning and transformers for aspect modeling are introduced.
Finally, we elaborate on the prediction and training of KCF-PLM.

\subsection{Overview}\label{subsec:overview}
Figure~\ref{fig:model} depicts the overall architecture of KCF-PLM.
For a given user-item pair, the input to KCF-PLM consists of three parts: (1) the text of the user reviews and the item reviews for the PLMs, (2) the aspects relevant to the user and the item for the transformer network, and (3) the KG for graph neural networks.
Within the model, the representations learned from the KG are used to enhance the user and item representations obtained by the PLMs and the input representations of the transformer.
And the enhanced user and item representations guide the attention-based combination of the aspect representations output by the transformer.
Based on the enhanced user and item representations, and the combined aspect representations, KCF-PLM generates the estimated rating score.
In what follows, we detail the key parts of the model.

\subsection{Aspect Extraction and Knowledge Graph}
\label{sec:kg}
\textbf{Aspect Extraction.} To utilize knowledge related to reviews for promoting rating prediction, we extract the fine-grained aspects from raw review texts.
The public available tool PyABSA\footnote{\url{https://github.com/yangheng95/PyABSA}} is adopted in this study and so that $\Set{A}_{u}$ and $\Set{A}_{i}$ are ready to be modeled.
It is worth noting that an aspect can occur multiple times in the reviews related to the given user or the given item.
It is intuitive that the aspects with larger occurrence counts might have more contributions to representations.
Therefore, we use this type of information in transformers for aspect modeling.

\begin{figure}[!t]
\centering
\includegraphics[width=.8\linewidth]{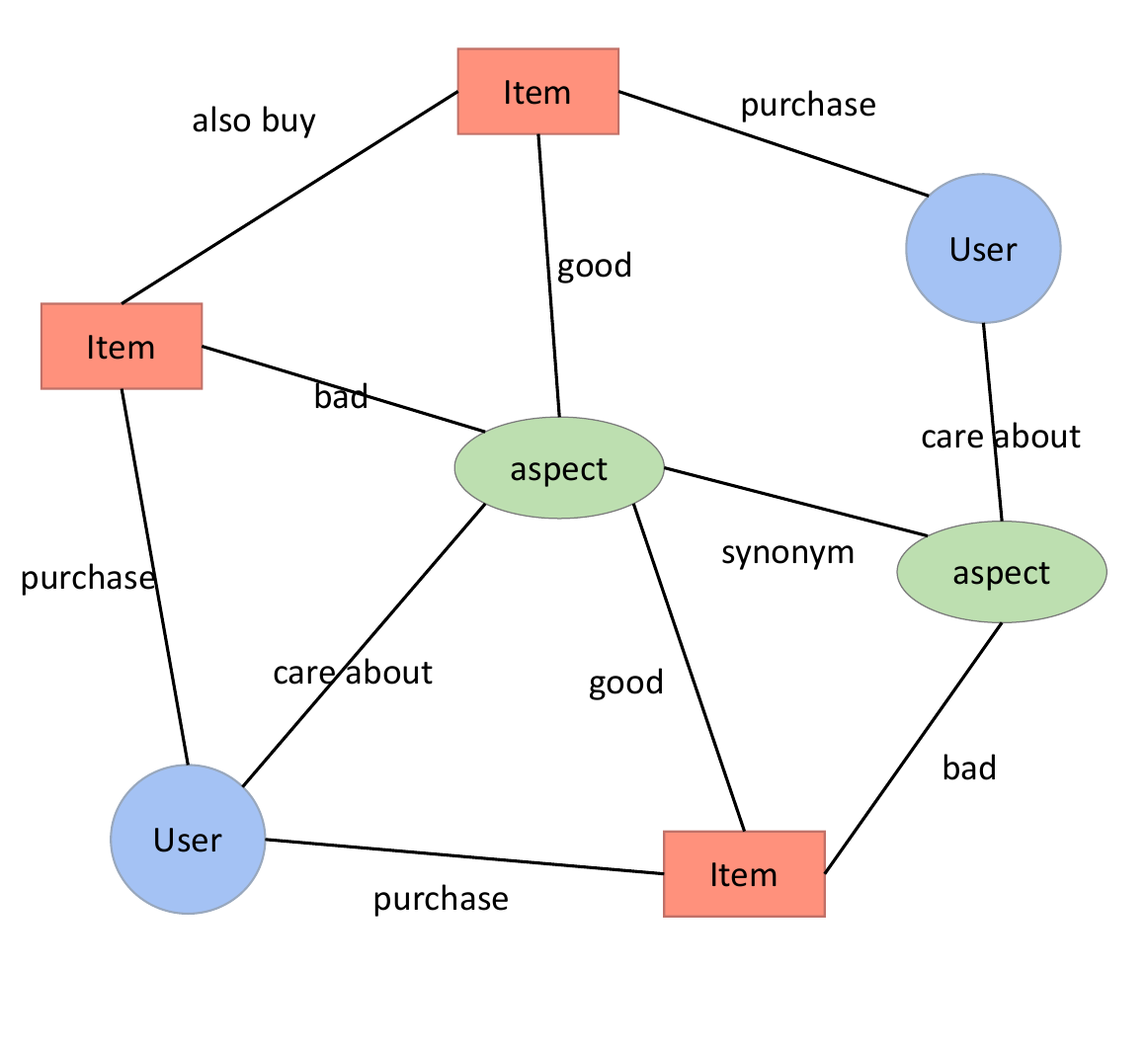}
\vspace{-2em}
\caption{Sketch of the constructed knowledge graph.}
\label{fig:KG}
\end{figure}

\textbf{KG Construction.}
The knowledge graph constructed in this paper is actually a heterogeneous graph $\boldsymbol{G}=(\Set{V},\Set{E})$, where $\Set{V}$ denotes the node set and $\Set{E}$ denotes the edge set.
A brief illustration of the constructed KG is shown in Figure~\ref{fig:KG}.
There are three types of nodes as follows: user nodes $\Set{V}^U$, item nodes $\Set{V}^I$, and aspect nodes $\Set{V}^A$.

For the edges $\Set{E}$ in the KG, we consider the following six types:
\begin{enumerate}
    \item user-item edge (purchase): denoting the action of purchase.
    \item item-aspect edge (good): denoting that a certain item is good w.r.t. a specific aspect.
    \item item-aspect edge (bad): denoting that a certain item is not good w.r.t. a specific aspect.
    \item user-aspect edge (care about): denoting that a certain user cares about a specific aspect.
    \item aspect-aspect edge (synonym): denoting that two aspects are synonymous.
    \item item-item edge (also purchase): denoting that there are common buyers of two items.
\end{enumerate}

The edges between nodes in the KG are created in the following manners.
1) For the type of user-item edge, the corresponding edges are built according to the available user-item interactions in training datasets.
2)-4) For the second to fourth edge types, we construct them based on the aspects and their corresponding sentiment polarities extracted by PyABSA.
It can extract the aspects and determine their corresponding sentiment polarities based on the raw texts. 
For an aspect, if it appears in the historical reviews of an item, we create the corresponding item-aspect edges.
The `good' edge and the `bad' item-aspect edge could appear at the same time since the item might receive both positive and negative sentiment polarities w.r.t. the aspect in its different reviews.
To tackle this, we associate the item-aspect edge (good) with the edge weight $w_e=\dfrac{n_P}{n_P+n_N}$ and item-aspect edge (bad) with the edge weight $w_e=\dfrac{n_N}{n_P+n_N}$, where $n_P$ represents the number of occurrences of an aspect with positive sentiment polarity in the item reviews and $n_N$ represents the number of negative occurrences.
Besides, the type of user-aspect edge is easily handled by checking which aspect occurs in the reviews of a certain user.
5) The fifth type of edge is used to capture the semantic similarity between any two aspects.
To realize this, we adopt Glove~\cite{glove} to train word embeddings and use them to measure the similarity of aspects.
If the cosine similarity is larger than the specified threshold (i.e., 0.8 used in the experiments), we connect the two aspects in the KG.
6) The type of item-item edge is determined based on whether an item appears in the ``also\_buy'' list of another item.
This type of edge is adopted to capture the direct correlations between items.

\textbf{KG Modeling.} This paper utilizes the constructed KG to perform node representation learning, which will be used to strengthen the representations related to the parts of the PLMs and the transformer.
Considering the heterogeneity of the KG, we adopt RGCN~\cite{rgcn} and further incorporate the edge weights into the node representation aggregation.
Note that our model is generalizable and other heterogeneous graph neural networks could also be leveraged.

For brevity, we use RGCN($\cdot$) to denote the graph representation learning, which is given as: 
\begin{equation}\label{eq:rgcn}
\Mat{V}^U,\Mat{V}^I,\Mat{V}^A=\mathrm{RGCN}(\boldsymbol{G})\,,
\end{equation}
where $\Mat{V}^U$, $\Mat{V}^I$, and $\Mat{V}^A$ are the user, item, and aspect representation matrices after graph propagation, respectively.
The node representations in the graph are updated as follows:
\begin{equation}\label{eq:rgcnw}
\Vector{v}_i^{(l+1)}=\sigma\left(\sum_{r\in R}\sum_{j\in N_i^r}\frac1{c_{i,r}}W_r^{(l)}\Vector{v}_j^{(l)}w_e+W_0^{(l)}\Vector{v}_i^{(l)}\right)\,,
\end{equation}
where $N_i^r$ denotes the neighbor node set of node $i$ connected by edge type $r \in R$ (defined in Section~\ref{sec:kg}). $\Vector{v}_i$ is the representation of node $i$ in the $l$-th layer of RGCN. $c_i^r$ is a problem-specific normalization constant that can be either learned or chosen in advance (e.g., $c_i^r = |N_i^r|$). $W_r^{l}$ is a linear transformation function, which transforms the neighbor nodes of the same type of edge using a parameter matrix.
$W_0^{l}$ transforms the representation of the same node in the previous layer to the current layer.
It is worth mentioning that the initial representations of all the nodes (e.g., $\Vector{v}_i^{(0)}$) are their ID-based representations (converted from
the original one-hot encodings of IDs).

To align the representation space of RGCN with other modules, we adopt a simple transformation by an MLP.
Taking user $u$ as an example, the transformed representation is given in the following way:
\begin{equation}
    \tilde{\Mat{V}}^U_u := \mathrm{MLP}(\Mat{V}^U_u)\,.
\end{equation}
Similarly, we can get transformed representations for items and aspects.
It is worth mentioning that the MLPs used by users, items, and aspects do not share the same weights to maintain discrimination between each other.

\subsection{PLMs for Representation Learning}
\label{sec:PLMS}
As introduced in Section~\ref{sec:problem}, for user $u$ and item $i$, we have the corresponding document $\Set{D}_{u}$ and $\Set{D}_{i}$ which are composed by their multiple reviews.
We choose BERT~\cite{bert} as the pre-trained language model to encode user and item representations, without loss of generality.

Consequently, we feed $\Set{D}_{u}$ and $\Set{D}_{i}$ into the BERT model, which is formulated as follows:
\begin{equation}\label{eq:bert}
\Mat{P}^U_u=\mathrm{U}\text{-}\mathrm{BERT}(\Set{D}_{u})\,,~~~\Mat{P}^I_i=\mathrm{I}\text{-}\mathrm{BERT}(\Set{D}_{i})\,,
\end{equation}
where $\Mat{P}^U_u$ and $\Mat{P}^I_i$ correspond to the representations of $[CLS]$ at the top layer of BERT.
We use $\mathrm{U}\text{-}\mathrm{BERT}$ and $\mathrm{I}\text{-}\mathrm{BERT}$ to clarify that the two BERT models share the same architecture but with different model parameter values.

Similar to getting the transformed representations from the KG, we also add an MLP layer to transform the representations from the unstructured review text and have the updated $\Mat{P}^U_u$ and $\Mat{P}^I_i$.
Up to now, we have two types of representations for both users and items: the representations from the KG and representations from the unstructured text.
Based on them, we can obtain the enhanced user and item representations as follows:
\begin{equation}\label{eq:enhanced-rep}
    \Mat{U}_u = \tilde{\Mat{V}}^U_u + \Mat{P}^U_u \,,  ~~~    \Mat{I}_i = \tilde{\Mat{V}}^I_i + \Mat{P}^I_i\,,
\end{equation}
where $\Mat{U}_u$ and $\Mat{I}_i$ are the enhanced representations of the user and the item, respectively.

\subsection{Transformer Networks for Aspect Modeling}
\label{sec:aspect}
KCF-PLM proposes to use transformer networks to model the interactions w.r.t. the aspects of a certain user-item pair.
As introduced in Section~\ref{sec:problem}, there are two aspect sets: $\Set{A}_{u}$ for user $u$ and $\Set{A}_{i}$ for item $i$.
Taking these aspects as input to the transformer, one straightforward strategy is to directly concatenate the two sets together.
Yet this strategy overlooks to distinguish the origin of the aspects.
To be concrete, some aspects might be shared by both user $u$ and item $i$, while other aspects might only belong to the user or the item. 
It is intuitive that the shared aspects might have different effects on the interaction modeling w.r.t. the user-item pair.

Based on the above consideration, we have the following representation types to characterize aspects as follows:
\begin{itemize}[leftmargin=*]
    \item KG-based aspect representations. They are learned by RGCN as shown in Eq.~\ref{eq:rgcn}, e.g., for aspect $a$, $\Vector{e}_a^{KG}=\Mat{V}_a^A$. 
    
    \item Source-based aspect representations. We define three types of sources, i.e., user, item, and user\&item. Each of them is assigned by a representation, denoted by $\Vector{e}_a^{SRC}$.\label{item:SRC_aspect}
\end{itemize}

To further capture the importance of each aspect, KCF-PLM uses the occurrence count of an aspect in the user and item review sets.
Suppose the count of aspect $a$ is $num_a$, then we compute the importance weight of the aspect by:
\begin{equation}\label{eq:aspect_weight}
w_a = 1 + (R-1)\cdot(\frac{2}{1 + exp(-num_a)} - 1)\,,
\end{equation}
where $R$ is the number of ratings (e.g., 5 in Amazon).
The underlying reasons for the design of the above equation are that: (1) A larger occurrence count of an aspect indicates it might be more important in forming representations. (2) The importance weight should be bounded. In other words, when $num_a=+\infty$, $w_a$ should not be $+\infty$. Through the above equation, the upper bound of importance weights is constrained to be R.
Consequently, the integrated representation of the aspect is calculated by:
\begin{equation}\label{eq:aspect_emb}
    \Vector{e}_a = w_a \cdot (\Vector{e}_a^{KG}  + \Vector{e}_a^{SRC})\,,
\end{equation}

\textbf{Transformer layer.} Given the above aspect representations, KCF-PLM feeds them to a stack of transformer layers for comprehensive interaction modeling, which is expressed by:
\begin{equation}\label{eq:transformer-aspect}
    \{\Vector{p}_1, \Vector{p}_2, \dots, \Vector{p}_{s_{ui}}\} = \text{Transformer}(\{\Vector{e}_1, \Vector{e}_2, \dots, \Vector{e}_{s_{ui}}\})\,.
\end{equation}
where $\{\Vector{p}_a\}_1^{s_{ui}}$ are the contextualized aspect representations output by transformers.

\textbf{User-item guided attention.}
KCF-PLM further employs user and item representations to guide attention-based fine-grained aspect modeling.
Specifically, the query in the attention computation is the concatenation of user and item representations, i.e., $\Vector{q}_{ui}=\Mat{U}_u\oplus\Mat{I}_i$.
Subsequently, the attention score $s_a$ for aspect $a$ is given by:
\begin{equation}\label{eq:attention}
    s_a = \Vector{w}^T \text{tanh}(\Mat{W}^T_q \Vector{q}_{ui} + \Mat{W}^T_p \Vector{p}_a + \Vector{b}_1) + b_2\,,
\end{equation}
where $\Mat{W}_q$ and $\Mat{W}_p$ are the matrix parameters, $\Vector{w}$ and $\Vector{b}_1$ are the vector parameters, and $b_2$ is a scalar parameter to be optimized.

Based on the attention scores, KCF-PLM aggregates the aspects into a wholistic representation as follows:
\begin{equation}\label{eq:aggregation}
    \Vector{p}_{ui}=\sum_{a=1}^{s_{ui}}\frac{\exp{(s_a)}}{\sum_{a'=1}^{s_{ui}}\exp{(s_{a'})}}\cdot\Vector{p}_a\,,
\end{equation}
where $\Vector{p}_{ui}$ is the aggregated representation for the user-item pair, which summarizes the fine-grained aspects.

\subsection{Prediction and Training}
Through the above computational procedures, we have $\Vector{q}_{ui}$ mainly from the pre-trained language models and $\Vector{p}_{ui}$ obtained by the transformer network.
To generate rating prediction, we first add these two representations, i.e., $\Vector{z}_{ui} = \Vector{q}_{ui} + \Vector{p}_{ui}$.
Then inspired by the study~\cite{deepconn}, we also use the Factorization Machine (FM)~\cite{Rendle12} to model the interactions between the latent dimension of representation $\Vector{z}_{ui}$ and obtain the rating score $ \hat{y}_{ui}$.
This computational formula is formulated as follows:
\begin{equation}\label{eq:fm-score}
    \hat{y}_{ui} = \text{FM}(\Vector{z}_{ui})\,.
\end{equation}
We adopt the Mean Squared Error (MSE) as the objective function to train KCF-PLM (including fine-tuning U-BERT and I-BERT), which is given by:
\begin{equation}\label{eq:loss-r}
    \Loss_r = \sum_{(u,i)} (\hat{y}_{ui} - y_{u,i})^2\,.
\end{equation}

To further benefit the representation learning of users, items, and aspects, we introduce another objective function w.r.t. KG.
This objective aims to make the representations obtained by RGCN retain their node-type information in KG.
Specifically, for a node $v$, its representation $\Vector{v}$ is gotten by Equation~\ref{eq:rgcn}.
Then its probability of belonging to each type is calculated as follows:
\begin{equation}\label{eq:node-prob}
\Vector{p}_v=[\Vector{p}_{v,1}, \Vector{p}_{v,2}, \Vector{p}_{v,3}]=\mathrm{softmax}(\mathrm{MLP}(\Vector{v}))\,,
\end{equation}
where $\Vector{p}_{v,1}$, $\Vector{p}_{v,2}$, and $\Vector{p}_{v,3}$ denote the probabilities of belonging to users, items, and aspects, respectively.
Given this, we define the following node-type cross-entropy loss (the effect of the loss is demonstrated in Table~\ref{tab:ablation_1}):
\begin{equation}\label{loss:node-type}
    \Loss_t = -\frac{1}{|\Set{V}|} \sum_{v=1}^{|\Set{V}|} \Vector{y}_v\log(\Vector{p}_v)\,,
\end{equation}
where $\Vector{y}_v$ denotes the one-hot representation of ground-truth node type for node $v$ and $|\Set{V}|$ is the number of nodes in the KG.

Finally, we utilize the node-type loss to complement the main loss, and the hybrid loss function for optimizing KCF-PLM is formulated as follows:
\begin{equation}\label{loss:hybrid}
    \Loss = \Loss_r + \alpha\Loss_t\,,
\end{equation}
where $\alpha$ is a hyperparameter to control the relative influence of $\Loss_t$ and is empirically set to 0.2.

\section{Experiments}
In this section, we conduct experiments to answer the following essential research questions:
\begin{itemize}
	\item[\textbf{\texttt{Q1}:}] How is the performance of KCF-PLM compared with other well-performed personalized review-based rating prediction models?
	\item[\textbf{\texttt{Q2}:}] What is the contribution of each key component in KCF-PLM?
\end{itemize}
Before delving into the experiments, we first provide the details about the datasets, baselines, and implementations in the following section.

\renewcommand{\arraystretch}{1.2}
\begin{table}[!t]
\centering
\caption{Statistics of the experimental datasets. We use \# to denote the number of instances.}
\resizebox{1.\linewidth}{!}{
\begin{tabular}{ |c |c |c |c |c |c |c |c|c|}
\hline
  & Clothing  & Movies & Toys &  Automotive & CellPhones & Yelp \\ 
\hline
 Label & 1-5  & 1-5 & 1-5 &  1-5 & 1-5 & 1-5 \\
\hline
 \#Train & 52453  & 44793 & 49755 &  41095 & 46466 & 32386\\
\hline
 \#Validation & 6512  & 5473 & 6528 &  4636 & 5934 & 2850 \\
\hline
 \#Test & 6512  & 5473 & 6528 &  4636 & 5934 & 2850 \\
\hline
\#Reviews & 65477 & 55739 & 62811 & 50367 & 58334 & 38086 \\
\hline
Density & 0.017\% & 0.020\% & 0.028\% & 0.019\% & 0.020\% & 0.017\% \\
\hline
\end{tabular}
}
\label{tab:datasets}
\end{table}

\begin{table}[!t]
\centering
\caption{Statistics of the constructed knowledge graph for each dataset.}
\resizebox{1.\linewidth}{!}{
\begin{tabular}{|c| c| c| c| c| c| c| c|c|}
\hline
 & Clothing  & Movies & Toys &  Automotive & CellPhones  & Yelp \\ 
 
 \hline
 \#Users &16624 &	15587 &	14437 &	15515 &	18006 & 13066 \\
\hline
 \#Items &	23216 &	17588 &	15651 &	16888 &	15832 & 14071 \\
\hline
 \#Aspects &	796 &	612 &	904 &	898 &	955 & 1124 \\
\hline
 \#Nodes &	40636 & 33787 & 30992 & 33301 & 34793 & 28261 \\
 \hline
 \#Edges &	680895 &	558791 &	675614 &	477855 &	727531 & 515213 \\
\hline
\end{tabular}
}
\label{tab:kg}
\end{table}

\subsection{Dataset}\label{sec:dataset}
We build the experimental datasets from two data sources.
The first one is Amazon and we choose the following five categories from the whole dataset~\cite{amazon-dataset} to build our Amazon datasets: (1) Clothing Shoes and Jewelry (abbreviated to Clothing), (2) Movies and TV (abbreviated to Movies), (3) Toys and Games (abbreviated to Toys), (4) Automotive, (5) Cell Phones and Accessories (abbreviated to Cell Phones).
The second one is Yelp and we use the version adopted in Kaggle competition\footnote{\url{https://www.kaggle.com/datasets/yelp-dataset/yelp-dataset/versions/6}}.

For each original dataset, we adopt the following preprocessing procedures: (1) removing users and items with less than 5 reviews; (2) using the open-source toolkit PyABSA for aspect extraction and aspect-based sentiment analysis; (3) constructing knowledge graphs by the approach introduced in Section~\ref{sec:kg}.

The summarized statistics of the experimental datasets are shown in Table~\ref{tab:datasets} and the details of the constructed knowledge graphs are presented in Table~\ref{tab:kg}.
To evaluate the performance on the datasets, we divide each dataset into the training, validation, and test sets by the ratio of 8 to 1 to 1.
The Mean Squared Error (MSE) is adopted as the evaluation metric commonly used by conventional studies.

\subsection{Parameter Setting}
To ensure a fair comparison, we tune all the baselines and determine their hyper-parameters based on their performance on the validation datasets, which is the same for tuning the proposed model KCF-PLM.
For KCF-PLM, we set the number of layers of RGCN and transformer networks as 1 and 4, respectively.
The maximal length of documents for users and items (e.g., $\Set{D}_{u}$ and $\Set{D}_{i}$) is set to 300.
The setting of the used BERT is by default, with 12 layers and 768-dimensional hidden representations.
The dimension of the KG nodes is set to 64.

We train the model with the Adam~\cite{adam} optimizer, with a learning rate of 6e-5 for 50 epochs in total.
We adopt early stopping to alleviate the overfitting issue.
The batch size is set to 12 to fully leverage the GPU memory.
All the source codes are implemented with the Pytorch library. 
The experiments are conducted on a Linux server with an RTX 3090 GPU.

\begin{table}[!t]
\caption{
Properties of different models. ``Review form'' indicates how the models organize historical reviews to constitute model input.
``User/Item ID'', ``KG'', ``Aspect'', and ``PLMs'' denote whether the models use the ID information, knowledge graphs, review aspects, and pre-trained language models, respectively.
}
\centering
\resizebox{1.\linewidth}{!}{
\begin{tabular}{| c | c | c | c | c | c |}
\hline
Method     & \makecell{Review form} & \makecell{User/Item ID} &	KG &	Aspect &	PLMs  \\ 
\hline
DeepCoNN   & Document&	 &  &	 &	   \\ 
\hline
NARRE     & Review &	\CheckmarkBold &  &	 &	   \\  
\hline
MPCN       & Review &	 & &	 &	 \\ 
\hline
AHN       & Review &	\CheckmarkBold &  &	  &	   \\  
\hline
AARM         & --- &	\CheckmarkBold &	 &	\CheckmarkBold &	   \\ 
\hline
SSG       & Document &	\CheckmarkBold &	\CheckmarkBold &	 &	  \\ \hline

NCEM        & Review &	\CheckmarkBold &	 &	 &	\CheckmarkBold   \\ \hline

DAML        & Document &	\CheckmarkBold &	 &	 &	  \\ 
\hline

CFRR        & --- &	\CheckmarkBold &	 &	\CheckmarkBold &	  \\ 
\hline

\makecell{KCF-PLM (Ours)} & Document &	\CheckmarkBold &	\CheckmarkBold &	\CheckmarkBold &	\CheckmarkBold  \\ 
\hline
\end{tabular}
}
\label{tab:model-features}
\end{table}

\subsection{Baseline}
We compare the proposed KCF-PLM with the following neuro-based rating prediction models proposed in recent years:

\begin{table*}[!t]
\caption{
Performance of our model and baselines on test sets. \textbf{MSE (lower is better)} is used for evaluation. The best performances are in \textbf{bold} and the second-best results are underlined. ``Error Reduction'' denotes the error percentage reduced by KCF-PLM over the best baseline in each dataset.}
\centering
\begin{tabular}{ c |c |c |c| c| c|c}
\toprule
Method  & Clothing  & Movies & Toys &  Automotive & CellPhones & Yelp  \\ 
\midrule

DeepCoNN  & 0.9826 & 0.9488 & 0.8221 & 1.1149 & 1.2769 & 1.5122  \\ 
\hline

NARRE    & 0.9515 &	0.9260 & 0.8099 & 1.1207 & 1.2703 &  1.5057 \\  
\hline

MPCN     & 1.0155 &	1.0359 & 0.8765 & 1.1475 & 1.2956 &  1.5591 \\ 
\hline

AHN      &  0.9890 & 0.9291 & 0.7979 & 1.1322 & 1.2859 &  1.4930\\  
\hline

AARM    & 0.9319 &	0.9380 & 0.7864 & 1.1145 & 1.2407 &  1.4667 \\ 
\hline 

SSG   &  0.9595 &	\underline{0.9185} & 0.7862 &	1.1121 & 1.2511  & 1.4578 \\  
\hline

NCEM   & 0.9294 & 0.9368 & 0.7774 & \underline{1.0916} &	1.2301 & 1.4336 \\  
\hline
DAML   & 0.9280  & 0.9337 & \underline{0.7693}  & 1.0934 & 1.2259  &1.4423   \\ 
\hline

CFRR & \underline{0.9165} & 0.9238 & 0.7789 & 1.0988 & \underline{1.2168} & \underline{1.4209}  \\
\hline

KCF-PLM (Ours) & \textbf{0.8740}  & \textbf{0.8712}  & \textbf{0.7478} & \textbf{1.0456} &  \textbf{1.1696} & \textbf{1.3336}  \\ 
\hline
Error Reduction& 4.64\% & 5.15\% & 2.79\% & 4.21\% & 3.88\% & 6.14\%       \\ 
\bottomrule
\end{tabular}
\label{tab:main-results}
\end{table*}

\begin{itemize}
\item \textbf{DeepCoNN~\cite{deepconn}}: Deep cooperative neural network is a pioneering neuro-based model that learns user preferences and item characteristics from historical review text for rating prediction.
CNNs are used as the backbone for text modeling.

\item \textbf{NARRE~\cite{narre}}: Neural attentional regression model encodes each review text separately and exploits the attention mechanism to learn the importance of different reviews according to the text representations as well as their corresponding user (item) IDs.

\item \textbf{MPCN~\cite{mpcn}}: Multi-pointer co-attention network models the interactions between user and item reviews at both review-level and word-level to search for important reviews and words. Instead of using soft attention weights to represent importance, it uses hard pointers for review selection.

\item \textbf{AHN~\cite{ahn}}: Asymmetrical hierarchical network argues that useful information from reviews is different for users and items, and therefore proposes to extract features from user and item reviews with asymmetric attentive modules.

\item \textbf{AARM~\cite{AARM}}: Attentive aspect-based recommendation model utilizes an attention mechanism to capture different attention of users on different aspects w.r.t. various items. Meanwhile, AARM models the interactions between synonymous aspects and similar aspects to compensate for the sparsity of aspects in the review texts.

\item \textbf{SSG~\cite{SSG}}: Set-Sequence-Graph is a multi-view model which organizes reviews as sets, sequences, and graphs views, respectively. 
SSG uses a tripartite encoder architecture to jointly capture long-term (sets), short-term (sequences), and collaboration (graphs) of user and item features for recommendation.

\item \textbf{NCEM~\cite{NCEM}}: Neural collaborative embedding model utilizes BERT to gain review-based representations in the first stage and then feed these representations to the rating prediction model in the second stage.

\item \textbf{DAML~\cite{daml}}: Dual attention mutual learning method utilizes local and mutual attention to extract user and item features from reviews, and performs higher-order nonlinear interaction between review-based and rating-based features.

\item \textbf{CFRR~\cite{CFRR}}: Counterfactual Review-based Recommendation employs counterfactual reasoning to analyze user reviews and gain a deeper understanding of user preferences. The system evaluates both consumed and non-consumed items, comparing actual user experiences with hypothetical scenarios where users might have interacted with different items. This approach enables the identification of patterns and preferences not captured by conventional methods.

\end{itemize}

The properties of the proposed model and the baseline models are summarized in Table~\ref{tab:model-features}.
Among them, KCF-PLM is the only one that simultaneously considers user/item IDs, KG, aspect, and PLMs.

\subsection{Model Comparison}
We run each experiment three times and report the average results of all the models in Table~\ref{tab:main-results}, from which we have the following observations.

\begin{itemize}[leftmargin=*]
\item Both DeepCoNN and MPCN only use the original review texts as input, but MPCN performs the worst in all baseline models. The reason is that DeepCoNN uses CNN to extract the information in user and item reviews, which makes it pay attention to the information contained in different reviews, while MPCN uses a hard pointer to select reviews and search for important reviews and words. This denotes that to extract accurate and comprehensive users' and items' features from history reviews, we need information from multiple reviews instead of only one or two reviews.

\item Both NARRE and AHN introduce the IDs of users and items as additional features, and the IDs also contain rich information.
Thus they have different degrees of improvement compared to DeepCoNN. 
NARRE performs better than AHN, which illustrates that using an attention mechanism to learn the importance of different reviews could better describe users and items.
AHN pays attention to the difference between user reviews and item reviews and uses an asymmetric attention mechanism to gain performance.

\item AARM abandons the use of original text data.
Instead, it uses ID information and aspect information to model users and items from coarse-grained and fine-grained, respectively. 
The overall performance is further improved, but other information in the text is ignored. 
The SSG model performs well on the Movies and TV datasets. 
It builds a knowledge graph for users and items.
However, the structure of its knowledge graph is relatively simple, with only a single relationship type.
Moreover, it is difficult to learn the sufficient social attributes of users and projects.

\item NCEM, DAML, and CFRR behave better than the other baselines in most cases.
For NCEM, its advantage is attributed to the use of pre-trained models to learn from the review text.
For DAML, its advantage lies in the use of hierarchical CNNs and attention mechanisms to enhance the original interaction between user and item reviews.
For CFRR, its advantage is using the counterfactual reasoning method like data augmentation.

\item KCF-PLM outperforms all the baseline methods on all six datasets significantly through t-test, with an error reduction of up to 6.14\% over the best baseline model.
This shows the applicability and effectiveness of the proposed model.
\end{itemize}

\begin{table*}[!t]
\caption{
Ablation study of KCF-PLM (MSE is used for evaluation).
}
\centering
\begin{tabular}{ |c|l| c| c| c| c| c|c|}
\hline
\multicolumn{2}{|c|}{Method}    & Clothing  & Movies & Toys &  Automotive & CellPhones & Yelp \\ 

\hline
\multicolumn{2}{|c|}{KCF-PLM} & \textbf{0.8740}  & \textbf{0.8712}  & \textbf{0.7478} & \textbf{1.0456} &  \textbf{1.1696} & \textbf{1.3336}  \\ 

\hline

\multirow{4}*{Components} & — w/o KG   &  0.8989 & 0.8976 & 0.7602 & 1.0657 & 1.2032 & 1.3796  \\ 

\cline{2-8}
& — w/o aspect transformer    & 0.9059 & 0.9017 & 0.7598 & 1.0732 & 1.1908 & 1.3948 \\ 

\cline{2-8}
& — w/o BERT (w/ DAML) & 0.8903 & 0.8922 & 0.7601 & 1.0688 & 1.1896 & 1.3937  \\ 
\hline
\multirow{5}*{Strategies} & — w/o node-type loss & 0.8861 & 0.8836 & 0.7529 & 1.0608 & 1.1886 & 1.3755 \\

\cline{2-8}
& — w/o aspect source emb  & 0.8901 & 0.8886 & 0.7595 &	1.0602 & 1.1876 & 1.3658 \\ 

\cline{2-8}
& — w/o weight   & 0.8816 & 0.8796 & 0.7533 & 1.0587 & 1.1856 & 1.3675   \\ 
\cline{2-8}
& — w/o attention   & 0.8917 & 0.8891 & 0.7612 & 1.0678 & 1.1887 & 1.3699  \\ 

\cline{2-8}
& — w/o fine-tuning  &  0.8813 & 0.8802 & 0.7536 &	1.0534 & 1.1708 & 1.3789  \\

\hline
\end{tabular}
\label{tab:ablation_1}
\end{table*}

\begin{table*}[!t]
\caption{
The train time cost (m, short for minutes) for reaching convergence and inference time cost (ms, short for milliseconds) for each data instance.}
\centering
\begin{tabular}{ |l|c| c| c| c| c| c|c|c|c| c| c|c|c|}
\hline

Dataset    & \multicolumn{2}{|c|}{Clothing}  & \multicolumn{2}{|c|}{Movies} & \multicolumn{2}{|c|}{Toys} &  \multicolumn{2}{|c|}{Automotive} & \multicolumn{2}{|c|}{CellPhones} & \multicolumn{2}{|c|}{Yelp} \\ 

\hline

Training (a) / Inference (b) & a &  b & a &  b & a  &  b & a &  b & a &  b & a &  b  \\

\hline

KCF-PLM & 242m & 7.60ms & 247m  & 7.49ms  & 259m  & 7.65ms & 330m & 7.37ms  & 216m & 7.24ms  & 105m & 7.36ms  \\

\hline

— w/o KG & 216m & 5.23ms & 192m &  5.25ms & 241m &  5.46ms & 325m &  5.21ms & 202m &  5.22ms & 101m &  5.22 ms  \\

\hline

— w/o aspect & 196m & 4.79ms & 175m &  4.77ms & 220m &  4.82ms & 281m &  4.77ms & 179m &  4.75 ms & 88m &  4.73 ms  \\

\hline
— w/o BERT (w/ DAML) & 123m & 2.58ms & 129m &  2.60ms & 127m &  2.62ms & 220m &  2.59ms & 133m &  2.49ms & 57m &  2.50 ms  \\

\hline

\end{tabular}
\label{tab:cost-time}
\end{table*}

\subsection{Ablation Study}
In this section, we perform an ablation study to show the effect of model components and computational strategies used in KCF-PLM.

\subsubsection{Effect of Model Components}\label{subsubsec:components}
The main components of KCF-PLM are the knowledge graph, transformer networks for aspects, and the pre-trained model BERT for review-based representations. To investigate the effectiveness of these components, we consider the following variants of KCF-PLM: (1) ``w/o KG'' denotes removing the knowledge graph (RGCN is also removed since it is purely used for the knowledge graph) from KCF-PLM. (2) ``w/o aspect transformer'' means removing the aspect modeling part, which is realized by deleting the transformer network from KCF-PLM.
(3) ``w/o BERT (w/DAML)'' represents replacing BERT by the backbone of DAML~\cite{daml} to extract user/item features from reviews. The choice of DAML is motivated by the fact that DAML shows good performance among the baselines in Table~\ref{tab:main-results} and outperforms other models like RNNs~\cite{ahn} commonly used in NLP. 
Thus it can be used as a strong competitor to verify the benefit of using PLMs in KCF-PLM.

We compare KCF-PLM with these alternatives.
The results are shown in the first part of Table~\ref{tab:ablation_1}, based on which we have the following observations.
Firstly, removing either the knowledge graph or the transformer network incurs notable performance degradation. 
These phenomena demonstrate the large performance gain brought by knowledge graph modeling w.r.t. different types of nodes and relations, and the transformer network for addressing aspects.
Moreover, removing both the aspect nodes from the KG and the transformer network for aspect modeling incurs a large performance degradation. 
Secondly, replacing BERT with the backbone of DAML also consistently reduces the performance, although the performance drop is not as large as the other two model variants.
This shows that pre-trained language models have an advantage over traditional models when modeling unstructured texts for the studied task.

\begin{figure*}[!t]
\centering
\begin{tikzpicture}
\begin{axis}
			[ybar=0.2cm,
			 symbolic x coords={Clothing,Movies,Toys,Automotive,CellPhones,Yelp},
			 legend style={at={(0.32,1)}},
              ylabel=MSE,
              width=16.5cm,
              height=6.1cm,
			 nodes near coords,
              nodes near coords style={font=\tiny},
			 enlarge x limits=0.1,
              xtick=data,
    ] 
 \addplot 
	coordinates {(Clothing,1.0824) (Movies,0.9632) (Toys,0.8631)(Automotive,1.1389) (CellPhones,1.4601) (Yelp,1.5666)};
 \addplot 
	coordinates {(Clothing,0.9665) (Movies,0.9204) (Toys,0.7782)(Automotive,1.0534) (CellPhones,1.2989) (Yelp,1.3267)};
 \addplot 
	coordinates {(Clothing,0.8667) (Movies,0.8562) (Toys,0.6456)(Automotive,1.0346) (CellPhones,1.2555) (Yelp,1.2890)};
 \addplot 
	coordinates {(Clothing,0.5956) (Movies,0.6838) (Toys,0.6308)(Automotive,0.8798) (CellPhones,0.9109) (Yelp,1.1024)};

\legend{$1 \le n < 12$,$12 \le n<24$,$24 \le n<36$,$n \ge 36$}
\end{axis}
\end{tikzpicture}
\caption{The influence of the aspect number on the rating prediction performance.}
\label{fig:num-aspects}
\end{figure*}
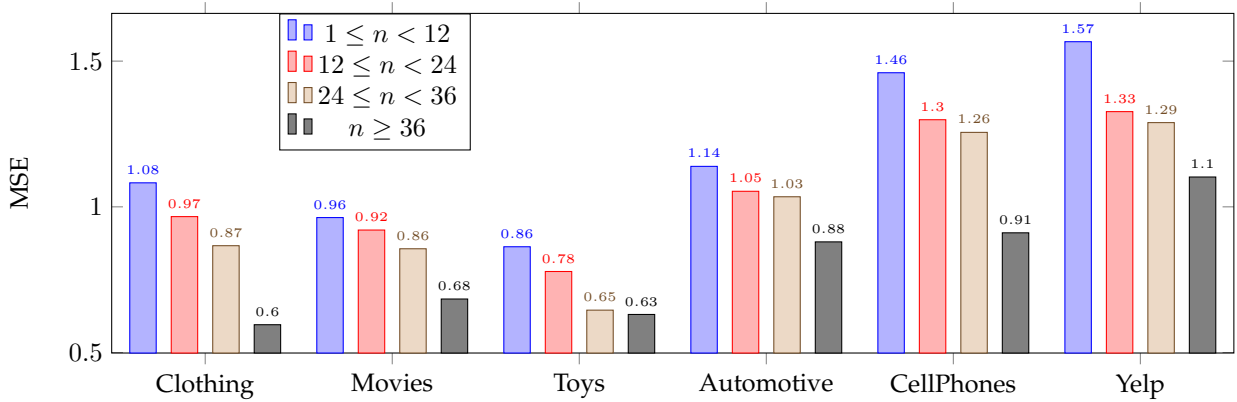

\subsubsection{Effect of Computational Strategies}
This part further performs an ablation study to validate the benefits of the computational strategies adopted by the proposed model.
To achieve this, we have the following variants by removing different strategies: (1) ``w/o node-type loss'' means removing the node-type loss by setting $\alpha$=$0$ in Equation~\ref{loss:hybrid}.
(2) ``w/o weight'' denotes all the aspects have the same weight (i.e., $w_a$=1 in Equation~\ref{eq:aspect_emb}). 
(3) ``w/o aspect source emb'' means removing the aspect source embedding. 
(4) ``w/o attention'' denotes attention is not used when calculating the output of the aspect sequence through transformer, but the mean value is taken instead. 
(5) ``w/o fine-tuning'' denotes not fine-tuning BERT when training KCF-PLM.
By observing the results in the second part of Table~\ref{tab:ablation_1}, we have the following findings.

\noindent$\bullet$ \textbf{Effect of Node-type Loss.} 
The results of ``w/o node-type loss'' indicates that utilizing node-type cross-entropy loss as a complementary objective function is beneficial for the rating prediction performance.

\noindent$\bullet$ \textbf{Effect of Aspect Weight.}
The results indicate that the aspect weight proposed in Equation~\ref{eq:aspect_weight} is profitable. 
This means aspects with more occurrence in the historical reviews of users and items tend to be more important.

\noindent$\bullet$ \textbf{Effect of Aspect Source Embedding.} The results show the source-based aspect representation $\Vector{e}_a^{SRC}$ is useful for improving the model performance.
This might be because aspects from different sources (i.e., user/item/both reviews) have different meanings. 
For example, an aspect appearing in user reviews reveals what the user cares about. 
By contrast, when it appears in item reviews, we could infer the property of the item.
This difference has an impact on the final prediction.

\noindent$\bullet$ \textbf{Effect of Attention.}
When the user-item guided attention mechanism is replaced by averaging the sequence output of aspects, the performance drop is observed in the table.
This is because users only pay attention to some aspects of interest when they buy products.
The attention mechanism helps the model to learn this information.

\noindent$\bullet$ \textbf{Effect of Fine-tuning.}
As the results are shown in the table, freezing the parameters of BERT in KCF-PLM incurs larger performance penalties than the other strategies.
Moreover, by comparing the results of ``w/o BERT (w/ DAML)'' in Table~\ref{tab:ablation_1} with the results of ``w/o fine-tuning'' in the table, we find that training the backbone of DAML along with the whole model also outperforms the way of freezing the parameters of BERT.
The above findings are due to the fact that freezing the parameters of BERT will lead to insufficient learning of user and item representations, which are very crucial for generating accurate personalized rating scores.

\begin{figure}[!t]
\centering
\begin{tikzpicture}
\pgfplotsset{every axis legend/.append style={
at={(0.5,1.03)},
anchor=south},every axis y label/.append style={at={(0.07,0.5)}}}
\begin{axis}[title=(a) Mean STD for ,xlabel=Iterations,
    ylabel=MSE,xtick ={2,4,6,8,10,12,14,16},legend columns=3,legend style={font=\tiny},
    xticklabel style={font=\footnotesize},
    height=6.125cm,
    width=7cm,
    ]
\addplot table [x=a, y=Type-1, col sep=comma] {mse.csv};
\addplot table [x=a, y=Type-2, col sep=comma] {mse.csv};
\addplot table [x=a, y=Type-3, col sep=comma] {mse.csv};
\addplot table [x=a, y=Type-4, col sep=comma] {mse.csv};
\addplot table [x=a, y=Type-5, col sep=comma] {mse.csv};
\addplot table [x=a, y=Type-6, col sep=comma] {mse.csv};

\legend{Clothing,Movies,Toys,Automotive,CellPhones,Yelp}
\end{axis}
\end{tikzpicture}
\caption{The change of the average MSE with the number of iterations in the training stage.}
\label{fig:mse_iter}
\end{figure}
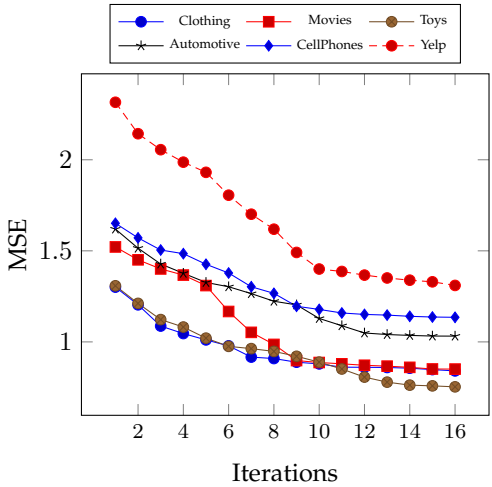

\subsection{Fine-grained Analysis of Test Data}\label{subsec:fine-grained}
We divide the data in the test sets into different buckets to investigate the effect of the number of input aspects on model performance. 
As shown in Figure~\ref{fig:num-aspects}, we set 4 intervals, i.e. [1,12), [12,24), [24,36), and [36,+$\infty$). 
And the buckets are built according to the interval in which the number of aspects is located. 
The results on the six datasets show that as the number of aspects increases, the rating prediction performance of the model is improved in most cases. 
This is because the more aspects contained in the reviews, the better they can reflect user preference and item properties, which directly leads to better user and item representations.

\subsection{Time Cost Analysis}\label{subsec:time-cost}
Table~\ref{tab:cost-time} shows the train and inference costs of the proposed model and its main variants --- ``w/o KG'', ``w/o aspect'', and ``w/o BERT (w/ DAML)''.
From a whole perspective, for the full model KCF-PLM, the training time costs are no more than 330m for all the six datasets and the inference costs are just over 7ms, which seem to be acceptable in real situations.
By further investigating the time costs of the three variants, we find that modeling KG by RGCN and modeling aspects by transformer network do not incur many time costs.
On the contrary, fine-tuning the pre-trained language model BERT in the training stage and using the pre-trained language model for inference dominate the time cost, which meets the expectation since the pre-trained language model is more complex than other model architectures.
We leave how to compress pre-trained language models in KCF-PLM for accelerating training and inference speed as future work.
Besides, we show the training curves w.r.t. MSE in Figure~\ref{fig:mse_iter}, where all the six datasets are involved. 
As can be seen, the training convergence is reached without using many iterations.

\begin{figure}[!t]
\centering
\begin{tikzpicture}
\pgfplotsset{every axis legend/.append style={
at={(0.5,1.03)},
anchor=south},every axis y label/.append style={at={(0.07,0.5)}}}
\begin{axis}[title=(a) Mean STD for ,xlabel=Num of RGCN layers,
    ylabel=MSE(Lower is better),xtick =data,legend columns=3,legend style={font=\tiny},font=\footnotesize,
    height=6.125cm,
    width=7cm,
    ]
\addplot table [x=a, y=Type-1, col sep=comma] {mydata.csv};
\addplot table [x=a, y=Type-2, col sep=comma] {mydata.csv};
\addplot table [x=a, y=Type-3, col sep=comma] {mydata.csv};
\addplot table [x=a, y=Type-4, col sep=comma] {mydata.csv};
\addplot table [x=a, y=Type-5, col sep=comma] {mydata.csv};
\addplot table [x=a, y=Type-6, col sep=comma] {mydata.csv};

\legend{Clothing,Movies,Toys,Automotive,CellPhones,Yelp}
\end{axis}

\end{tikzpicture}
\caption{Influence of the layer number of RGCN on the rating prediction performance.}
\label{fig:RGCN}
\end{figure}
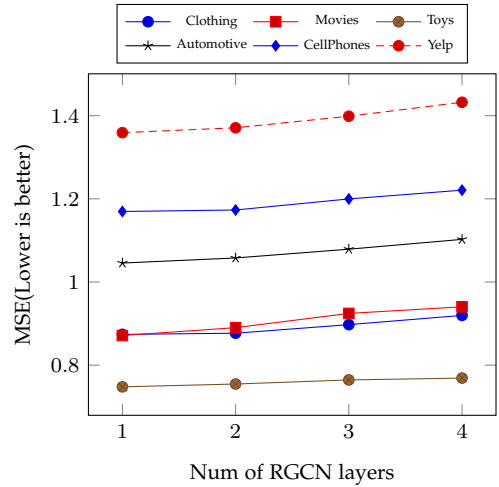

\subsection{Hyperparameter Analysis}
This part mainly analyzes the impact of the layer number of RGCN layers on the rating prediction performance.
As the results are shown in Figure~\ref{fig:RGCN}, we can observe that our model tends to perform better when the knowledge graph is trained with one RGCN layer.
This might be because we have already modeled the interaction between nodes explicitly with the six types of edges.
More R-GCN layers for information transfer might cause the model to be overfitting.

\subsection{Case Study}
\begin{figure}[t]
\centering 

\includegraphics[scale=0.38,trim=100 110 100 100,clip]{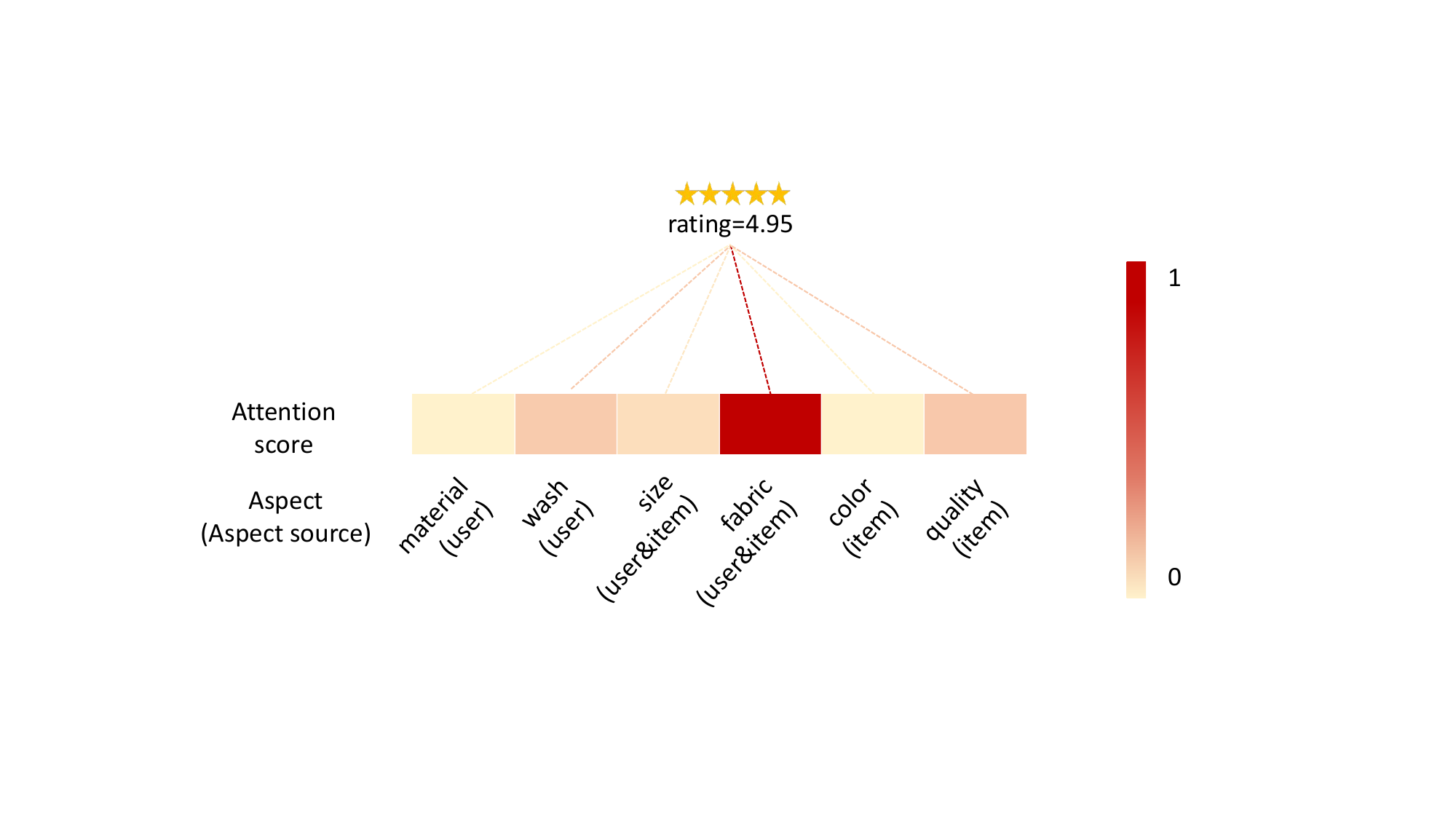}
\caption{A real example for visualizing the attention scores on aspects.}
\label{real-case}
\end{figure}

In this section, we conduct a case study to explore whether KCF-PLM can capture fine-grained aspect information. As shown in Figure~\ref{real-case}, among all the aspects of the input, ``fabric'' has the largest attention score. And we notice that it exists on both the user side and the item side, which is in line with our expectations. 
In addition, in conjunction with Figure~\ref{fig:running-case}, we retrieve the original test data and find that the review of the user 'A3HPFC4WMB2LZD' to the item 'B00FLXEP8O' (noting that it does not appear in the historical reviews of the user and the item) also mentions ``fabric''. This shows that the fine-grained modeling of KCF-PLM is indeed effective.

\section{Conclusion}
In this paper, we study the personalized review-based rating prediction task.
The existing studies in this regard mainly suffer from two issues:
(1) overlooking to model the fine-grained aspects in reviews and unstructured review text to complement each other;
(2) barely considering leveraging the power of pre-trained language models for this task.
To tackle the above issues, we propose a novel model named KCF-PLM, which incorporates a knowledge graph and extracted aspects into the model effectively.
A transformer network is developed to learn the interactions of structured review aspects and pre-trained language models are leveraged to learn the representations of unstructured review text in an end-to-end fashion.
The extensive experiments demonstrate the superiority of the proposed KCF-PLM model.

\ifCLASSOPTIONcaptionsoff
  \newpage
\fi



%
\bibliographystyle{ieeetr}
\bibliography{reference}

%




\begin{IEEEbiography}[{\includegraphics[width=1in,height=1.25in,clip,keepaspectratio]{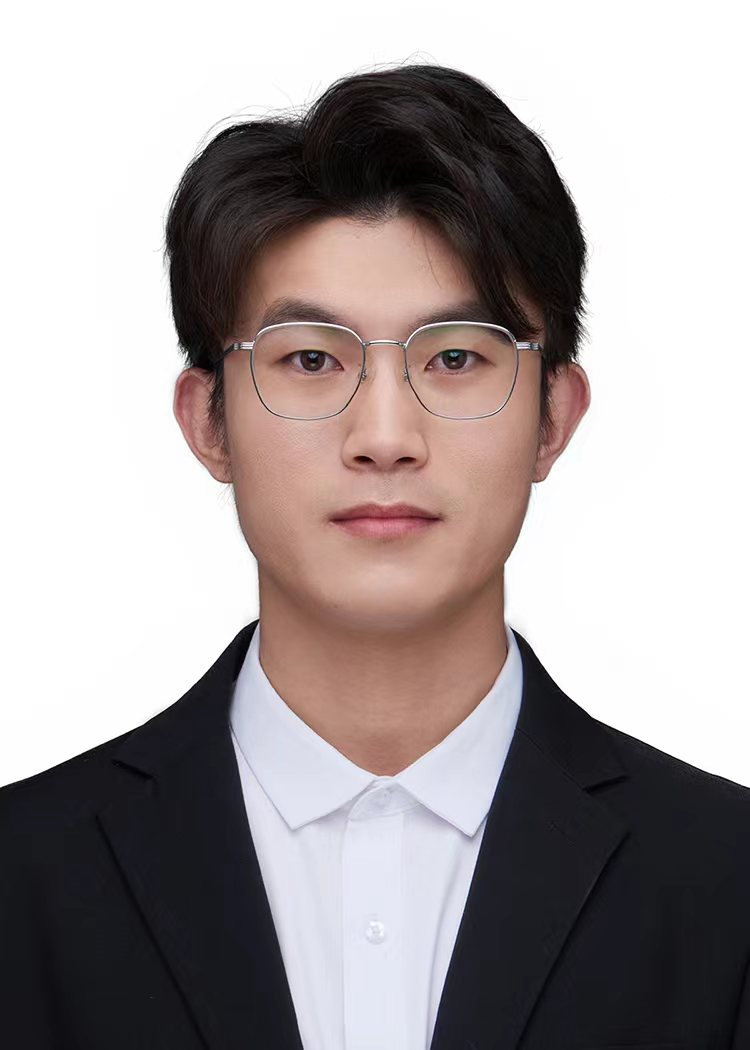}}]{Quanxiu Wang} received the B.S. degree in Computer Science and Engineering from Shandong University of Science and Technology, China, in 2020. He is currently pursuing the master's degree with the Department of Computer Science and Technology, East China Normal University, China. His research interests include recommendation systems, natural language processing and pre-trained language models.
\end{IEEEbiography}

\begin{IEEEbiography}[{\includegraphics[width=1in,height=1.25in,clip,keepaspectratio]{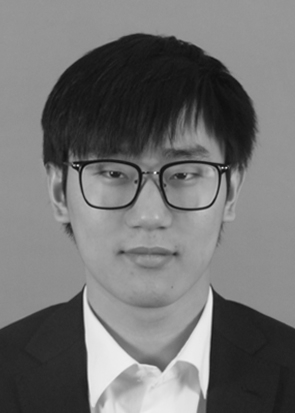}}]{Xinlei Cao} received the B.S. degree in Software Engineering from East China Normal University, China, in 2019. He is currently pursuing the master's degree with the Department of Computer Science and Technology, East China Normal University, China. His research interests include natural language processing, pre-trained language models and information retrieval.
\end{IEEEbiography}

\begin{IEEEbiography}[{\includegraphics[width=1in,height=1.25in,clip,keepaspectratio]{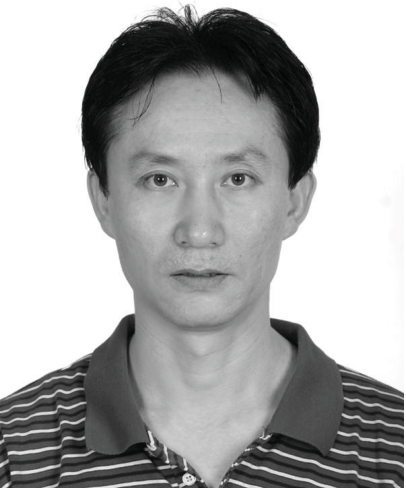}}]{Jianyong Wang} received the PhD degree in computer science from the Institute of Computing Technology, Chinese Academy of Sciences, in 1999. He is currently a professor in the Department of Computer Science and Technology, Tsinghua University, Beijing, China, and also with the Jiangsu Collaborative Innovation Center for Language Ability, Jiangsu Normal University, Xuzhou, China. His research interests mainly include data mining and Web information management. He has co-authored more than 60 papers in some leading international conferences and some top international journals. He is serving or has served as a PC member for some leading international conferences, such as SIGKDD, VLDB, ICDE, WWW, and an associate editor of the IEEE Transactions on Knowledge and Data Engineering and the ACM Transactions on Knowledge Discovery from Data. He is a fellow of the IEEE.
\end{IEEEbiography}

\begin{IEEEbiography}[{\includegraphics[width=1in,height=1.25in,clip,keepaspectratio]{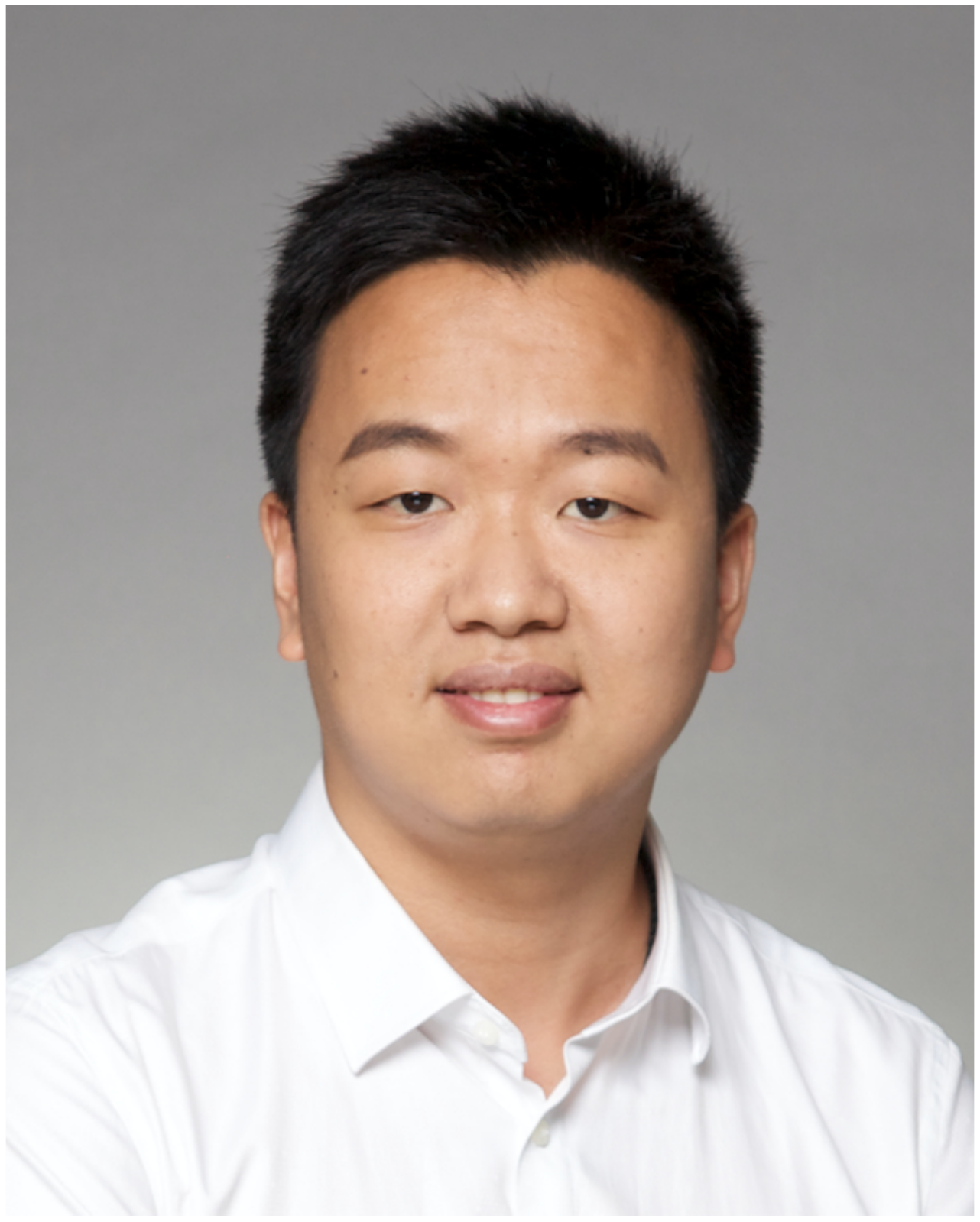}}]{Wei Zhang} received his Ph.D. degree in computer science and technology from Tsinghua university, Beijing, China, in 2016. He is currently a professor in the School of Computer Science and Technology, East China Normal University, Shanghai, China.
His research interests mainly include user data mining and machine learning applications.
He is a senior member of China Computer Federation.
\end{IEEEbiography}





\end{document}